\documentclass[aoas,preprint]{imsart}

\RequirePackage[OT1]{fontenc}
\RequirePackage{natbib}
\RequirePackage[colorlinks,citecolor=blue,urlcolor=blue]{hyperref}
\usepackage{amsmath, amstext, amsfonts, lscape, tikz}
\usepackage{graphicx,psfrag,epsf}
\usepackage{enumerate}
\usepackage{float}
\usepackage{subfig}
\usepackage{url}
\usepackage{todonotes}
\usepackage{multirow,rotating}
\usepackage[framemethod=TikZ]{mdframed}
\usepackage[textheight=622pt,textwidth=448pt,centering,headheight=50pt,headsep=12pt,footskip=12pt]{geometry}

\setattribute{journal}{name}{}
\arxiv{arXiv:1602.07933}

\startlocaldefs
\numberwithin{equation}{section}

\endlocaldefs

\begin{document}

\begin{frontmatter}
\title{Bootstrap Inference When Using Multiple Imputation}
\runtitle{Bootstrap Inference When Using Multiple Imputation}

\begin{aug}
\author{\fnms{Michael} \snm{Schomaker}\ead[label=e1]{michael.schomaker@uct.ac.za}}
\and\author{\fnms{Christian} \snm{Heumann}\ead[label=e2]{chris@stat.uni-muenchen.de}}

\affiliation{University of Cape Town\thanksmark{m1} and Ludwig-Maximilians Universit{\"a}t M{\"u}nchen\thanksmark{m2}}

\address{Michael Schomaker\\
Centre for Infectious Disease Epidemiology \& Research\\
University of Cape Town\\
Cape Town, South Africa\\
\printead{e1}\\
\phantom{E-mail:\ michael.schomaker@uct.ac.za}}

\address{Christian Heumann\\
Institut f{\"u}r Statistik\\
Ludwig-Maximilians Universit{\"a}t M{\"u}nchen\\
M{\"u}nchen, Germany\\
\printead{e2}\\
\phantom{E-mail:\ chris@stat.uni-muenchen.de}}
\end{aug}

\begin{abstract}
Many modern estimators require bootstrapping to calculate confidence intervals because either no analytic standard error is available or the distribution of the parameter of interest is non-symmetric. It remains however unclear how to obtain valid bootstrap inference when dealing with multiple imputation to address missing data. We present four methods which are intuitively appealing, easy to implement, and combine bootstrap estimation with multiple imputation. We show that three of the four approaches yield valid inference, but that the performance of the methods varies with respect to the number of imputed data sets and the extent of missingness.  Simulation studies reveal the behavior of our approaches in finite samples. A topical analysis from HIV treatment research, which determines the optimal timing of antiretroviral treatment initiation in young children, demonstrates the practical implications of the four methods in a sophisticated and realistic setting. This analysis suffers from missing data and uses the $g$-formula for inference, a method for which no standard errors are available.\end{abstract}


\end{frontmatter}

\begin{mdframed}[backgroundcolor=red!10, linecolor=black!50]
The published version of this working paper can be cited as follows:\\[0.25cm]
Schomaker, M., Heumann, C.\\
\textit{Bootstrap Inference When Using Multiple Imputation}\\
Statistics in Medicine, 37(14):2252-2266\\
\url{http://dx.doi.org/10.1002/sim.7654}
\end{mdframed}

\section{Introduction}

Multiple imputation (MI) is a popular method to address missing data. Based on assumptions about the data distribution (and the mechanism which gives rise to the missing data) missing values can be imputed by means of draws from the posterior predictive distribution of the unobserved data given the observed data. This procedure is repeated to create $M$ imputed data sets, the analysis is then conducted on each of these data sets and the $M$ results ($M$ point and $M$ variance estimates) are combined by a set of simple rules \cite{Rubin:1996}.

During the last 30 years a lot of progress has been made to make MI useable for different settings: implementations are available in several software packages \cite{Horton:2007, Honaker:2011, vanBuuren:2011, Royston:2011}, review articles provide guidance to deal with practical challenges \cite{White:2011, Sterne:2009, Graham:2009}, non-normal --possibly categorical-- variables can often successfully be imputed \cite{Schafer:2002, Honaker:2011, White:2011}, useful diagnostic tools have been suggested \cite{Honaker:2011, Eddings:2012}, and first attempts to address longitudinal data and other complicated data structures have been made \cite{Honaker:2010, vanBuuren:2011}.

While both opportunities and challenges of multiple imputation are discussed in the literature, we believe an important consideration regarding the inference after imputation has been neglected so far: if there is no analytic or no ideal solution to obtain standard errors for the parameters of the analysis model, and nonparametric bootstrap estimation is used to estimate them, it is unclear how to obtain valid inference -- in particular how to obtain appropriate confidence intervals. Moreover, bootstrap estimation is also often used when a parameter's distribution is assumed to be non-normal and bootstrap inference with missing data is then not clear either. As we will explain below, many modern statistical concepts, often applied to inform policy guidelines or enhance practical developments, rely on bootstrap estimation. It is therefore necessary to have guidance for bootstrap estimation for multiply imputed data.

In general, one can distinguish between two approaches for bootstrap inference when using multiple imputation: with the first approach, $M$ imputed datsets are created and bootstrap estimation is applied to each of them; or, alternatively, $B$ bootstrap samples of the original data set (including missing values) are drawn and in each of these samples the data are multiply imputed. For the former approach one could use bootstrapping to estimate the standard error in each imputed data set and apply the standard MI combining rules; alternatively, the $B \times M$ estimates could be pooled and 95\% confidence intervals could be calculated based on the 2.5$^{th}$ and 97.5$^{th}$ percentiles of the respective empirical distribution. For the latter approach either multiple imputation combining rules can be applied to the imputed data of each bootstrap sample to obtain $B$ point estimates which in turn may be used to construct confidence intervals; or the $B \times M$ estimates of the pooled data are used for interval estimation.

To the best of our knowledge, the consequences of using the above approaches have not been studied in the literature before. The use of the bootstrap in the context of missing data has often been viewed as a frequentist alternative to multiple imputation \cite{Efron:1994}, or an option to obtain confidence intervals after single imputation \cite{Shao:1996}. The bootstrap can also be used to create multiple imputations \cite{Little:2002}. However, none of these studies have addressed the construction of bootstrap confidence intervals when data needs to be multiply imputed because of missing data. As emphasized above, this is however of particularly great importance when standard errors of the analysis model cannot be calculated easily, for example for causal inference estimators (e.g. the g-formula).

It is not surprising that the bootstrap has nevertheless been combined with multiple imputation for particular analyses. Multiple imputation of bootstrap samples has been implemented in \cite{Briggs:2006, Schomaker:2014, Schomaker:2015, Worthington:2014}, whereas bootstrapping the imputed data sets was preferred by \cite{Wu:2013, Baneshi:2012, Heymans:2007}. Other work doesn't offer all details of the implementation \cite{Chaffee:2014}. All these analyses give however little justification for the chosen method and for some analyses important details on how the confidence intervals were calculated are missing; it seems that pragmatic reasons as well as computational efficiency typically guide the choice of the approach. None of the studies offer a statistical discussion of the chosen method.

The present article demonstrates the implications of different methods which combine bootstrap inference with multiple imputation. It is novel in that it introduces four different, intuitively appealing, bootstrap confidence intervals for data which require multiple imputation, illustrates their intrinsic features, and argues which of them is to be preferred.

Section \ref{sec:motivation} introduces our motivating analysis of causal inference in HIV research. The different methodological approaches are described in detail in Section \ref{sec:methods} and are evaluated by means of both numerical investigations (Section \ref{sec:simulations}) and theoretical considerations (Section \ref{sec:theory}). The implications of the different approaches are further emphasized in the data analysis of Section \ref{sec:data}. We conclude in Section \ref{sec:conclusion}.

\clearpage

\section{Motivation}\label{sec:motivation}
During the last decade the World Health Organization (WHO) updated their recommendations on the use of antiretroviral drugs for treating and preventing HIV infection several times. In the past, antiretroviral treatment (ART) was only given to a child if his/her measurements of CD4 lymphocytes fell below a critical value or if a clinically severe event (such as tuberculosis or persistent diarrhoea) occurred. Based on both increased knowledge from trials and causal modeling studies, as well as pragmatic and programmatic considerations, these criteria have been gradually expanded to allow earlier treatment initiation in children: in 2013 it was suggested that all children who present under the age of 5 are treated immediately, while for older children CD4-based criteria still existed. By the end of 2015 WHO decided to recommend immediate treatment initiation in all children and adults. ART has shown to be effective and to reduce mortality in infants and adults \cite{Westreich:2012, Edmonds:2011, Violari:2008}, but concerns remain due to a potentially increased risk of toxicities, early development of drug resistance, and limited future options for people who fail treatment.

It remains therefore important to investigate the effect of different treatment initiation rules on mortality, morbidity and child development outcomes; however given the shift in ART guidelines towards earlier treatment initiation it is not ethically possible anymore to conduct a trial which answers this question in detail. Thus, observational data can be used to obtain the relevant estimates. Methods such as inverse probability weighting of marginal structural models, the g-computation formula, and targeted maximum likelihood estimation can be used to obtain estimates in complicated longitudinal settings where time-varying confounders affected by prior treatment are present --- such as, for example, CD4 count which influences both the probability of ART initiation and outcome measures \cite{Daniel:2013, Petersen:2014}.

In situations where treatment rules are dynamic, i.e. where they are based on a time-varying variable such as CD4 lymphocyte count, the g-computation formula \cite{Robins:1986} is \textit{the} intuitive method to use. It is computationally intensive and allows the comparison of outcomes for different treatment options; confidence intervals are typically based on non-parametric bootstrap estimation. However, in resource limited settings data may be missing for administrative, logistic, and clerical reasons, as well as due to loss to follow-up and missed clinic visits. Depending on the underlying assumptions about the reasons for missing data, this problem can either be addressed by the g-formula directly or by using multiple imputation. However, it is not immediately clear how to combine multiple imputation with bootstrap estimation too obtain valid confidence intervals.

\section{Methodological Framework}\label{sec:methods}
Let $\mathcal{D}$ be a $n \times (p+1)$ data matrix consisting of an outcome variable $\mathbf{y} = (y_1,\ldots,y_n)'$ and covariates $\mathbf{X_j} = (X_{1j},\ldots,X_{nj})'$, $j=1,\ldots,p$. The $1\times p$ vector $\mathbf{x}_i=(x_{i1},\ldots,x_{ip})$ contains the $i^{th}$ observation of each of the $p$ covariates and $\mathbf{X}=({\mathbf{x}_1}',\ldots,{\mathbf{x}_n}')'$ is the matrix of all covariates. Suppose we are interested in estimating $\theta = (\theta_1,\ldots,\theta_k)'$, $k \geq 1$, which may be a regression coefficient, an odds ratio, a factor loading, or an counterfactual outcome. If some data are missing, making the data matrix to consist of both observed and missing values, $\mathcal{D}=\{\mathcal{D}^{\text{obs}},\mathcal{D}^{\text{mis}}\}$, and the missingness mechanism is ignorable, valid inference for $\theta$ can be obtained using multiple imputation. Following Rubin \cite{Rubin:1996} we regard valid inference to mean that the point estimate $\hat{\theta}$ for $\theta$ is approximately unbiased and that interval estimates are randomization valid in the sense that actual interval coverage equals the nominal interval coverage.

Under multiple imputation $M$ augmented sets of data are generated, and the imputations (which replace the missing values) are based on draws from the posterior predictive distribution of the missing data given the observed data $p(\mathcal{D}^{\text{mis}}|\mathcal{D}^{\text{obs}})$  $=$  $\int p(\mathcal{D}^{\text{mis}}|\mathcal{D}^{\text{obs}};\vartheta)$ $p(\vartheta|\mathcal{D}^{\text{obs}}) \ d\vartheta$, or an approximation thereof. The point estimate for $\theta$ is
\begin{eqnarray}\label{formula_MI_point}
\hat{\bar\theta}_{\text{MI}} = \frac{1}{M} \sum_{m=1}^{M} \hat{\theta}_{m}
\end{eqnarray}
where $ \hat{\theta}_{m}$ refers to the estimate of $\theta$ in the $\text{m}^{th}$ imputed set of data $\mathcal{D}^{(m)}$, $m=1,\ldots,M$.  Variance estimates can be obtained using the between imputation covariance $\hat{V}= (M-1)^{-1} \sum_{m} (\hat{\theta}_{m}-\hat{\bar\theta}_{\text{MI}}) (\hat{\theta}_{m}-\hat{\bar\theta}_{\text{MI}})^{'}$ and the average within imputation covariance $\widehat{W}= M^{-1}$ $\sum_m \widehat{\text{Cov}}(\hat{\theta}_{m})$:
\begin{eqnarray}\label{formula_MI_var}
\widehat{{\text{Cov}}}(\hat{\bar\theta}_{\text{MI}}) &=& \widehat{W} + \frac{M+1}{M}\hat{V} = \frac{1}{M} \sum_{m=1}^{M} \widehat{\text{Cov}}(\hat{\theta}_{m}) \\&&+ \frac{M+1}{M(M-1)} \sum_{m=1}^{M} (\hat{\theta}_{m}-\hat{\bar\theta}_{\text{MI}}) (\hat{\theta}_{m}-\hat{\bar\theta}_{\text{MI}})^{'}\,.\nonumber
\end{eqnarray}
For the scalar case this equates to
\begin{eqnarray}
\widehat{{\text{Var}}}(\hat{\bar\theta}_{\text{MI}}) &=& \frac{1}{M} \sum_{m=1}^{M} \widehat{\text{Var}}(\hat{\theta}_{m}) + \frac{M+1}{M(M-1)} \sum_{m=1}^{M} (\hat{\theta}_{m}-\hat{\bar\theta}_{\text{MI}})^2\,.\nonumber
\end{eqnarray}
To construct confidence intervals for $\hat{{{\bar\theta}}}_{\text{MI}}$ in the scalar case, it may be assumed that  $\widehat{{\text{Var}}}(\hat{\bar\theta}_{\text{MI}})^{-\frac{1}{2}}(\hat{{\bar\theta}}_{\text{MI}}-\theta)$ follows a $t_{R}$-distribution with approximately $R=(M-1)[1+\{M\hat{W}/(M+1)\hat{V}\}]^2$ degrees of freedom \cite{Rubin:1986}, though there are alternative approximations, especially for small samples \cite{Lipsitz:2002}. Note that for reliable variance estimation $M$ should not be too small; see White et al. \cite{White:2011} for some rules of thumb.

Consider the situation where there is no analytic or no ideal solution to estimate $\text{Cov}(\hat{\theta}_{m})$, for example when estimating the treatment effect in the presence of time-varying confounders affected by prior treatment using g-methods \cite{Robins:2009, Daniel:2013}. If there are no missing data, bootstrap percentile confidence intervals may offer a solution: based on $B$ bootstrap samples $\mathcal{D}^{\ast}_b$, $b=1,\ldots,B$, we obtain $B$ point estimates $\hat{\theta}^{\ast}_b$. Consider the ordered set of estimates $\Theta^{\ast}_B = \{\hat{\theta}^{\ast}_{(b)}; b=1,\ldots,B \}$, where $\hat{\theta}^{\ast}_{(1)} < \hat{\theta}^{\ast}_{(2)} < \ldots < \hat{\theta}^{\ast}_{(B)}$; the bootstrap $1-2\alpha$\% confidence interval for $\theta$ is then defined as
\begin{eqnarray*}
[\hat{\theta}_{\text{lower}}; \hat{\theta}_{\text{upper}}] & = & [\hat{\theta}^{\ast,\alpha}; \hat{\theta}^{\ast,1-\alpha}]
\end{eqnarray*}
where $\hat{\theta}^{\ast,\alpha}$ denotes the $\alpha$-percentile of the ordered bootstrap estimates $\Theta^{\ast}_B$. However, in the presence of missing data the construction of confidence intervals is not immediately clear as $\hat{\theta}$ corresponds to $M$ estimates $\hat{\theta}_1,\ldots,\hat{\theta}_M$, i.e. $\hat{\theta}_m$ is the point estimate calculated from the $m^{th}$ imputed data set. It seems intuitive to consider the following four approaches:
\begin{itemize}
\item \textbf{Method 1, MI Boot (pooled sample [PS]):} Multiple imputation is utilized for the data set $\mathcal{D}=\{\mathcal{D}^{\text{obs}},\mathcal{D}^{\text{mis}}\}$. For each of the $M$ imputed data sets $\mathcal{D}_m$, $B$ bootstrap samples are drawn which yields $M \times B$ data sets $\mathcal{D}^{\ast}_{m,b}; b=1,\ldots,B; m=1,\ldots,M$. In each of these data sets the quantity of interest is estimated, that is $\hat{\theta}^{\ast}_{m,b}$. The pooled sample of ordered estimates $\Theta^{\ast}_{\text{MIBP}} = \{\hat{\theta}^{\ast}_{(m,b)}; b=1,\ldots,B; m=1,\ldots,M\}$ is used to  construct the $1-2\alpha$\% confidence interval for $\theta$:
    \begin{eqnarray}
    [\hat{\theta}_{\text{lower}}; \hat{\theta}_{\text{upper}}]_{\text{MIBP}} & = & [\hat{\theta}^{\ast,\alpha}_{\text{MIBP}} \hat{\theta}^{\ast,1-\alpha}_{\text{MIBP}}]
    \end{eqnarray}
    where $\hat{\theta}^{\ast,\alpha}_{\text{MIBP}}$  is the $\alpha$-percentile of the ordered bootstrap estimates $\Theta^{\ast}_{\text{MIBP}}$.\\
\item \textbf{Method 2, MI Boot:} Multiple imputation is utilized for the data set $\mathcal{D}=\{\mathcal{D}^{\text{obs}},\mathcal{D}^{\text{mis}}\}$. For each of the $M$ imputed data sets $\mathcal{D}_m$, $B$ bootstrap samples are drawn which yields $M \times B$ data sets $\mathcal{D}^{\ast}_{m,b}; b=1,\ldots,B;$ $m=1,\ldots,M$. The bootstrap samples are used to estimate the standard error of (each scalar component of) $\hat{\theta}_m$ in each imputed data set respectively, i.e. $\widehat{\text{Var}}(\hat{\theta}_m) = (B-1)^{-1} \sum_b (\hat{\theta}_{m,b} - \hat{\bar{\theta}}_{m})^2$ with $\hat{\bar{\theta}}_{m} = B^{-1}\sum_b \hat{\theta}_{m,b}$. This results in $M$ point estimates (calculated from the imputed, but not yet bootstrapped data), and $M$ standard errors (calculated from the respective bootstrap samples). More generally, $\text{Cov}(\hat{\theta}_{m})$ can be estimated in each imputed data set using bootstrapping, thus allowing the use of (\ref{formula_MI_var}) and standard multiple imputation confidence interval construction, possibly based on a $t_R$-distribution.\\
\item \textbf{Method 3, Boot MI (pooled sample [PS]):} $B$ bootstrap samples $\mathcal{D}^{\ast}_b$ (including missing data) are drawn and multiple imputation is utilized in each bootstrap sample. Therefore, there are $B \times M$ imputed data sets $\mathcal{D}^{\ast}_{b,1},\ldots,\mathcal{D}^{\ast}_{b,M}$ which can be used to obtain the corresponding point estimates $\hat{\theta}^{\ast}_{b,m}$. The set of the pooled ordered estimates $\Theta^{\ast}_{\text{BMIP}} = \{\hat{\theta}^{\ast}_{(b,m)}; b=1,\ldots,B; m=1,\ldots,M\}$ can then be used to construct the $1-2\alpha$\% confidence interval for $\theta$:
    \begin{eqnarray}
    [\hat{\theta}_{\text{lower}}; \hat{\theta}_{\text{upper}}]_{\text{BMIP}} & = & [\hat{\theta}^{\ast,\alpha}_{\text{BMIP}}; \hat{\theta}^{\ast,1-\alpha}_{\text{BMIP}}]
    \end{eqnarray}
    where $\hat{\theta}^{\ast,\alpha}_{\text{BMIP}}$  is the $\alpha$-percentile of the ordered bootstrap estimates $\Theta^{\ast}_{\text{BMIP}}$.\\
\item \textbf{Method 4, Boot MI:} $B$ bootstrap samples $\mathcal{D}^{\ast}_b$ (including missing data) are drawn, and each of them is imputed $M$ times. Therefore, there are $M$ imputed data sets, $\mathcal{D}^{\ast}_{b,1}, \ldots,$ $\mathcal{D}^{\ast}_{b,M}$, which are associated with each bootstrap sample $\mathcal{D}^{\ast}_b$. They can be used to obtain the corresponding point estimates $\hat{\theta}^{\ast}_{b,m}$. Thus, applying (\ref{formula_MI_point}) to the estimates of each bootstrap sample yields $B$ point estimates $\hat{\bar{\theta}}^{\ast}_b = M^{-1} \sum_m \hat{\theta}^{\ast}_{b,m}$ for $\theta$. The set of ordered estimates $\Theta^{\ast}_{\text{BMI}} = \{\hat{\theta}^{\ast}_{(b)}; b=1,\ldots,B\}$ can then be used to construct the $1-2\alpha$\% confidence interval for $\theta$:
    \begin{eqnarray}
    [\hat{\theta}_{\text{lower}}; \hat{\theta}_{\text{upper}}]_{\text{BMI}} & = & [\hat{\theta}^{\ast,\alpha}_{\text{BMI}}; \hat{\theta}^{\ast,1-\alpha}_{\text{BMI}}]
    \end{eqnarray}
    where $\hat{\theta}^{\ast,\alpha}_{\text{BMI}}$  is the $\alpha$-percentile of the ordered bootstrap estimates $\Theta^{\ast}_{\text{BMI}}$.\\
\end{itemize}

While all of the methods described above are straightforward to implement it is unclear if they yield valid inference, i.e. if the actual interval coverage level equals the nominal coverage level. Before we delve into some theoretical and practical considerations we expose some of the intrinsic features of the different interval estimates using Monte Carlo simulations.

\section{Simulation Studies}\label{sec:simulations}
To study the performance of the methods introduced above we consider four simulation settings: a simple one, to ensure that these comparisons are not complicated by the simulation setup; a more complicated one, to study the four methods under a more sophisticated variable dependence structure; a survival analysis setting to allow comparisons beyond a linear regression setup; and a complex longitudinal setting where time-dependent confounding (affected by prior treatment) is present, to allow comparisons to our data analysis in Section \ref{sec:data}.\\

\noindent\textit{Setting 1:}
We simulate a normally distributed variable $X_1$ with mean $0$ and variance $1$. We then define $\mu_y=0 + 0.4X_1$ and $\theta = \beta_{\text{true}}=(0,0.4)'$.
The outcome is generated from $N(\mu_y,2)$ and the analysis model of interest is the linear model. Values of $X_1$ are defined to be missing with probability
\begin{eqnarray*}
\pi_{X_1}(y) &=& 1 - \frac{1}{(0.25y)^2+1}\,.
\end{eqnarray*}
With this, about 16\% of values of $X_1$ were missing (at random).\\

\noindent\textit{Setting 2:} The observations for 6 variables are generated using the following normal and binomial distributions: $\mathbf{X}_1 \sim \text{N}(0,1)$, $\mathbf{X}_2 \sim \text{N}(0,1)$, $\mathbf{X}_3 \sim \text{N}(0,1)$, $\mathbf{X}_4 \sim \text{B}(0.5)$,  $\mathbf{X}_5 \sim \text{B}(0.7)$, and $\mathbf{X}_6 \sim \text{B}(0.3)$. To model the dependency between the covariates we use a Clayton Copula \cite{Yan:2007} with a copula parameter of $1$ which indicates moderate correlations among the covariates. We then define $\mu_y= 3 - 2X_1 + 3X_3 - 4X_5$ and $\theta = \beta_{\text{true}}=(3,-2,0,3,0,-4,0)'$.
The outcome is generated from $N(\mu_y,2)$ and the analysis model of interest is the linear model. Values of $X_1$ and $X_3$ are defined to be missing (at random) with probabilities
\begin{eqnarray*}
\pi_{X_1}(y) = 1 - \frac{1}{(ay)^2+1}\,,\quad\quad
\pi_{X_3}(X_4) = 1 - \frac{1}{ b X_4^3+1.05}\,.
\end{eqnarray*}
where $a$ and $b$ equate to $0.75$ and $0.25$ in a low missingness setting (and to $0.4$ and $2.5$ in a high missingness setting).
This yields about 6\% and 14\% (45\% and 38\%) of missing values for $X_1$ and $X_3$ respectively.\\

\noindent\textit{Setting 3:} This setting is inspired by the analysis and data in Schomaker et al. \cite{Schomaker:2015b}. We simulate $\mathbf{X}_1 \sim \text{logN}(4.286,1.086)$ and $\mathbf{X}_2 \sim \text{logN}(10.76,1.8086)$. Again, the dependency of the variables is modeled with a Clayton copula with a copula parameter of 1. Survival times $y$ are simulated from $-\log(U)/h_0\{\exp(X\beta)\}$ where $U$ is drawn from a distribution that is uniform on the interval $[0,1]$, $h_0=0.1$, and the linear predictor $X\beta$ is defined  as $-0.3 \ln X_1 + 0.3 \log_{10} X_2$. Therefore, $\beta_{\text{true}}=(-0.3,0.3)'$. Censoring times are simulated as $-\log(U)/0.2$. The observed survival time $T$ is thus $\min(y,C)$. Values of $X_1$ are defined to be missing based on the following function:
\begin{eqnarray*}
\pi_{X_1}(T) = 1 - \frac{1}{(0.075T)^2+1}\,.
\end{eqnarray*}
This yields about 8\% of missing values.\\

\noindent\textit{Setting 4:} This setting is inspired by our data analysis from Section \ref{sec:data}. We generate longitudinal data ($t=0,1,\ldots,12$) for 3 time-dependent confounders ($\mathbf{L_t}=\{L^1_t,L^2_t,L^3_t\}$), an outcome $(Y_t)$, an intervention $(A_t)$, as well as baseline data for 7 variables, using structural equation models \cite{Sofrygin:2016}. The data generating mechanism and the motivation thereof is described in Appendix \ref{sec:appendix_data_generating}. In this simulation we are interested in an counterfactual outcome $Y_t$ which would have been observed under 2 different intervention rules $\bar{d}^j$, $j=1,2$, which assign treatment ($A_t$) always or never. We denote these target quantities as $\psi_1$ and $\psi_2$ and their true values are $-1.03$ and $-2.45$ respectively. They can be estimated using the sequential g-formula, with bootstrap confidence intervals, see Appendix \ref{sec:appendix_g_formula} for more details.

Values of $L^1_t$, $L^2_t$, $L^3_t$, $Y_t$ are set to be missing based on a MAR process as described in Appendix \ref{sec:appendix_data_generating}.  This yields about 10\%, 31\%, 22\% and 44\% of missing baseline values, and 10\%, 1\%, 1\%, and 2\% of missing follow-up values.\\

In all 4 settings multiple imputation is utilized with \texttt{Amelia II} under a joint modeling approach, see Honaker et al. \cite{Honaker:2011} and Section 6 for details. In settings 1-3 the probability of a missing observation depends on the outcome. One would therefore expect parameter estimates in a regression model of a complete case analysis to be biased, but estimates following multiple imputation to be approximately unbiased \cite{Robins:1994, Little:1992}.

We estimate the confidence intervals for the parameters of interest using the aforementioned four approaches, as well as using the analytic standard errors obtained from the linear model and the Cox proportional hazards model (method ``no bootstrap'') for the first three settings. The ``no bootstrap'' method serves therefore as a gold standard and reference for the other methods. We generate $n=1000$ observations, $B=200$ bootstrap samples, and $M=10$ imputations. Based on $\mathcal{R}=1000$ simulation runs we evaluate the coverage probability and median width of the respective confidence intervals.\\

\noindent{\textit{Results}}: The computation time for Boot MI was always greater than for MI Boot, for example by a factor of $13$ in the first simulation setting and by a factor of $1.3$ in the fourth setting.

In all settings the point estimates for $\beta$ were approximately unbiased.

Table 1 summarizes the main results of the simulations. Using no bootstrapping yields estimated coverage probabilities of about 95\%, for all parameters and settings, as one would expect.

\begin{table}[h!]
\caption{\label{tab:sim_1_2}Results of the simulation studies: estimated coverage probability (top), median confidence intervals width (middle), and standard errors for different methods (bottom). The bottom panel lists standard errors estimated from the $1000$ point estimates of the simulation (``simulated'') and the mean estimated standard error across the simulated data sets, for both the analytical standard error (``no bootstrap'') and the bootstrap standard error (``MI Boot''). All results are based on $200$ bootstrap samples and $10$ imputations.}
\centering
{{
\fbox{%
\begin{tabular}{ccl|c|p{0.05\textwidth}p{0.05\textwidth}p{0.05\textwidth}p{0.05\textwidth}p{0.05\textwidth}p{0.05\textwidth}|p{0.05\textwidth}p{0.05\textwidth}}
&& \multicolumn{1}{c}{Method} & \multicolumn{1}{c}{Setting 1} & \multicolumn{6}{c}{Setting 2 (low missingness)}  & \multicolumn{2}{c}{Setting 3} \\
\hline
&&\multicolumn{1}{c}{}&\multicolumn{1}{c}{$\beta_1$}&$\beta_1$&$\beta_2$&$\beta_3$&$\beta_4$&$\beta_5$&$\beta_6$&\multicolumn{1}{c}{$\beta_1$}&$\beta_2$\\
\hline
\parbox[t]{2mm}{\multirow{5}{*}{\rotatebox[origin=c]{90}{Coverage}}}& \parbox[t]{2mm}{\multirow{5}{*}{\rotatebox[origin=c]{90}{Probability}}} & 1) MI Boot (PS)        &93\%&95\%&95\%&94\%&94\%&95\%&95\%&95\%&95\%\\
&& 2) MI Boot           &95\%&95\%&95\%&95\%&94\%&95\%&95\%&95\%&95\%\\
&& 3) Boot MI (PS)  &97\% &96\%&95\%&96\%&95\%&96\%&96\%&96\%&96\%\\
&& 4) Boot MI           &94\% &94\%&94\%&94\%&94\%&94\%&94\%&94\%&94\%\\
&& 5) no bootstrap      &95\% &95\%&95\%&95\%&95\%&95\%&96\%&95\%&95\%\\
\hline
\parbox[t]{2mm}{\multirow{5}{*}{\rotatebox[origin=c]{90}{Median}}}& \parbox[t]{2mm}{\multirow{5}{*}{\rotatebox[origin=c]{90}{CI Width}}} & 1) MI Boot (PS)             &0.30& 0.33 & 0.33 & 0.33 & 0.60 & 0.68 & 0.62 &0.25 & 0.31\\
&& 2) MI Boot           &0.31& 0.34 & 0.34 & 0.34 & 0.61 & 0.69 & 0.63 &0.26 & 0.31\\
&& 3) Boot MI (PS)  &0.35& 0.36 & 0.35 & 0.35 & 0.64 & 0.72 & 0.66 &0.26 & 0.31\\
&& 4) Boot MI           &0.30& 0.33 & 0.33 & 0.33 & 0.60 & 0.67 & 0.62 &0.24 & 0.30\\
&& 5) no bootstrap      &0.31& 0.34 & 0.34 & 0.34 & 0.61 & 0.69 & 0.63 &0.25 & 0.31\\
\hline
\parbox[t]{2mm}{\multirow{3}{*}{\rotatebox[origin=c]{90}{Std.}}}& \parbox[t]{2mm}{\multirow{3}{*}{\rotatebox[origin=c]{90}{Error}}}
&  simulated         &0.08& 0.09 & 0.09 & 0.09 & 0.16 & 0.18 & 0.16 & 0.06 & 0.08\\
&& no bootstrap      &0.08& 0.09 & 0.09 & 0.09 & 0.16 & 0.18 & 0.16 & 0.06 & 0.08\\
&& MI Boot           &0.08& 0.09 & 0.09 & 0.09 & 0.16 & 0.18 & 0.16 & 0.07 & 0.08\\
\hline
\multicolumn{11}{c}{}\\
&& \multicolumn{1}{c}{Method} & \multicolumn{1}{c}{} & \multicolumn{6}{c}{Setting 2 (high missingness)}  & \multicolumn{2}{c}{Setting 4} \\
\hline
&&\multicolumn{1}{c}{}&\multicolumn{1}{c}{}&$\beta_1$&$\beta_2$&$\beta_3$&$\beta_4$&$\beta_5$&$\beta_6$&\multicolumn{1}{c}{$\psi_1$}&$\psi_2$\\
\hline
\parbox[t]{2mm}{\multirow{5}{*}{\rotatebox[origin=c]{90}{Coverage}}}& \parbox[t]{2mm}{\multirow{5}{*}{\rotatebox[origin=c]{90}{Probability}}} & 1) MI Boot (PS)        &&89\%&91\%&92\%&91\%&92\%&92\%&94\%&94\%\\
&& 2) MI Boot           &&91\%&93\%&94\%&94\%&94\%&94\%&94\%&94\%\\
&& 3) Boot MI (PS)  &&96\%&97\%&98\%&98\%&97\%&98\%&94\%&94\%\\
&& 4) Boot MI           &&90\%&93\%&93\%&95\%&94\%&94\%&94\%&92\%\\
&& 5) no bootstrap      &&91\%&93\%&94\%&94\%&94\%&94\%&--&--\\
\hline
\parbox[t]{2mm}{\multirow{5}{*}{\rotatebox[origin=c]{90}{Median}}}& \parbox[t]{2mm}{\multirow{5}{*}{\rotatebox[origin=c]{90}{CI Width}}} & 1) MI Boot (PS)             && 0.44 & 0.40 & 0.44 & 0.79 & 0.87 & 0.78 &0.20 &0.21 \\
&& 2) MI Boot           && 0.48 & 0.44 & 0.49 & 0.87 & 0.95 & 0.86 &0.20 &0.22 \\
&& 3) Boot MI (PS)  && 0.58 & 0.51 & 0.59 & 1.03 & 1.12 & 1.01 &0.21 &0.23 \\
&& 4) Boot MI           && 0.47 & 0.43 & 0.47 & 0.84 & 0.92 & 0.82 &0.20 &0.22 \\ && 5) no bootstrap      && 0.48 & 0.44 & 0.49 & 0.87 & 0.95 & 0.86 &-- &-- \\
\hline
\parbox[t]{2mm}{\multirow{3}{*}{\rotatebox[origin=c]{90}{Std.}}}& \parbox[t]{2mm}{\multirow{3}{*}{\rotatebox[origin=c]{90}{Error}}}
&  simulated         && 0.12 & 0.11 & 0.12 & 0.22 & 0.24 & 0.21 &--  &-- \\
&& no bootstrap      && 0.12 & 0.12 & 0.12 & 0.22 & 0.24 & 0.22 &--  &-- \\
&& MI Boot           && 0.12 & 0.11 & 0.12 & 0.22 & 0.24 & 0.21 &--  &-- \\
\hline

\end{tabular}
}}}
\end{table}

Bootstrapping the imputed data (MI Boot, MI Boot [PS]) yields estimated coverage probabilities of about 95\% and confidence interval widths which are similar to each other, except for the high missingness setting of simulation 2. The standard errors for each component of $\beta$ as simulated in the 1000 simulation runs were almost identical to the mean estimated standard errors under MI Boot, which suggests good standard error estimation of the latter approach. In the first simulation setting the coverage of MI Boot pooled is a bit too low for $M=10$ (93\%), but is closer to 95\% if M is large ($M=20$, Figure 1).

Imputing the bootstrapped data (Boot MI, Boot MI [PS]) led to overall good results with coverage probabilities close to the nominal level, except for the high missingness setting of simulation 2; however, using the pooled samples led to somewhat higher coverage probabilities and the interval widths were slightly different from the estimates obtained under no bootstrapping.

Figure 1 shows the coverage probability of the interval estimates for $\beta_1$ in the first simulation setting given the number of imputations.

\begin{figure}[h!]
\begin{center}
\includegraphics[scale=0.4]{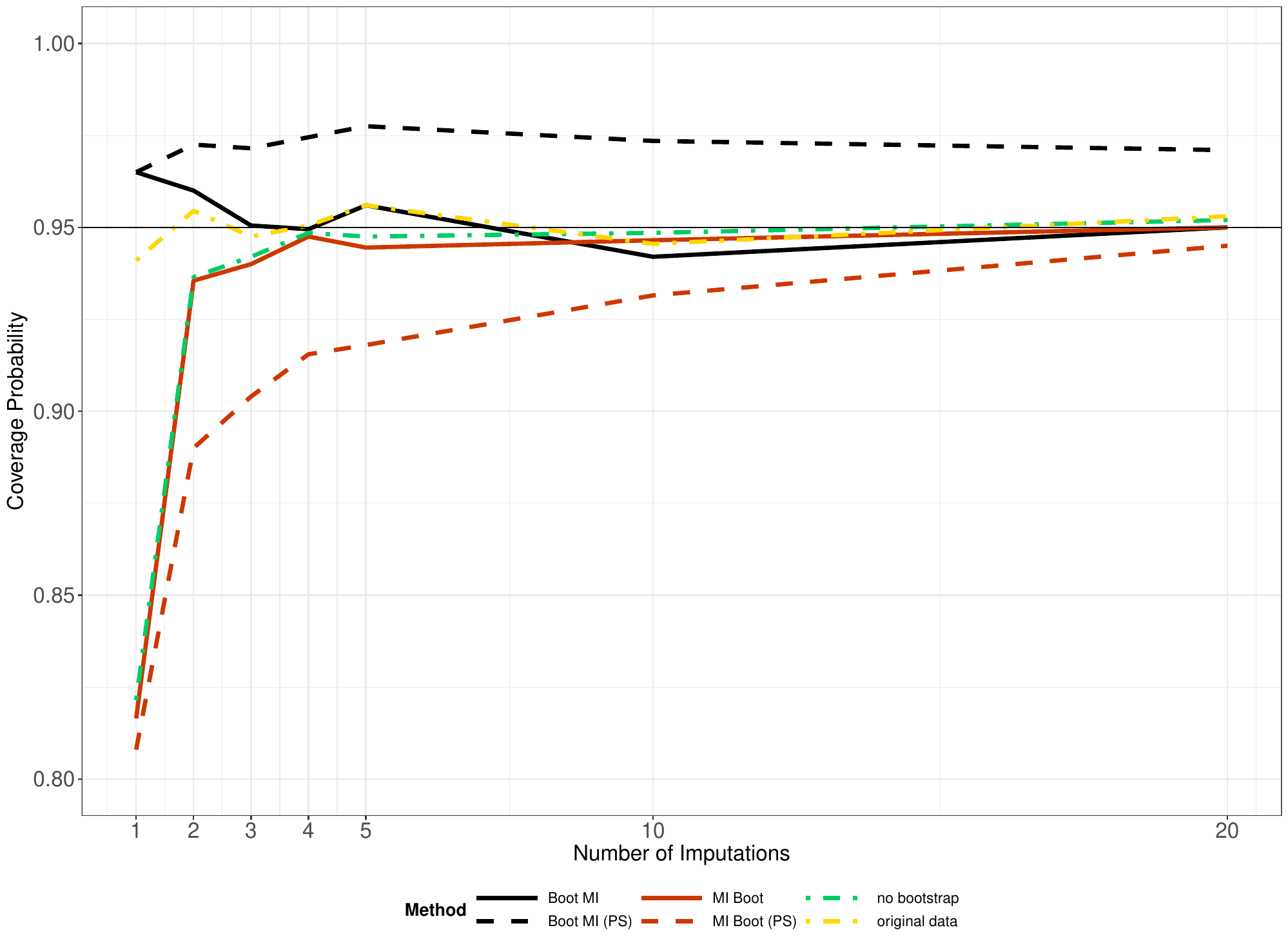}
\caption{\label{figure:sim1} Coverage probability of the interval estimates for $\beta_1$ in the first simulation setting dependent on the number of imputations. Results related to the complete simulated data, i.e. before missing data are generated, are labelled ``original data''.}
\label{fig1}
\end{center}
\end{figure}

\begin{figure}[h!]
\begin{center}
\includegraphics[scale=0.65]{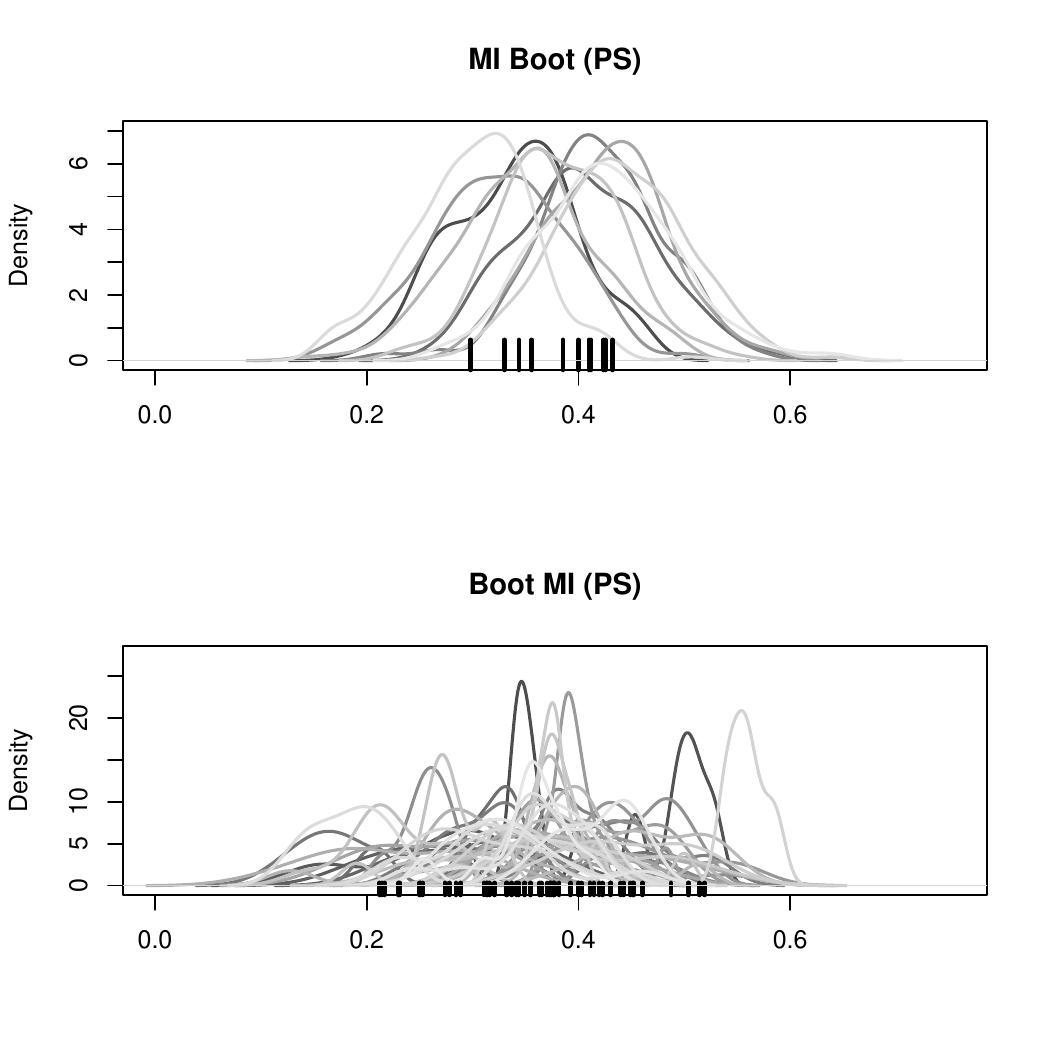}
\caption{\label{figure:sim2} Estimate of $\beta_1$ in the first simulation setting, for a random simulation run: distribution of `MI Boot (pooled)' for each imputed dataset (top) and distribution of `Boot MI (PS)' for 50 random bootstrap samples (PS). Point estimates are marked by the black tick marks on the x-axis.}
\label{fig2}
\end{center}
\end{figure}

As predicted by MI theory, using multiple imputation needs generally a reasonable amount of imputed data sets to perform well -- no matter whether bootstrapping is used for standard error estimation or not (MI Boot, no bootstrap). Boot MI may perform well even for $M<5$, but the pooled approach has a tendency towards coverage probabilities $>95\%$. For $M=1$ the estimated coverage probability of Boot MI is too large in the above setting.

Figure 2 offers more insight into the behaviour of `Boot MI (PS)' and `MI Boot (PS)' by visualizing both the bootstrap distributions in each imputed data set (method MI Boot [PS]) as well as the distribution of the estimators in each bootstrap sample (method Boot MI [PS]): one can see the slightly wider spectrum of values in the distributions related to `Boot MI (PS)' explaining the somewhat larger confidence interval in the first simulation setting.

More explanations and interpretations of the above results are given in Section \ref{sec:theory}.

\section{Data Analysis}\label{sec:data}
Consider the motivating question introduced in Section \ref{sec:motivation}. We are interested in comparing mortality with respect to different antiretroviral treatment strategies in children between 1 and 5 years of age living with HIV. We use data from two big HIV treatment cohort collaborations (IeDEA-SA, \cite{Egger:2012}; IeDEA-WA, \cite{Ekouevi:2011}) and evaluate mortality for $3$ years of follow-up. Our analysis builds on a recently published analysis by Schomaker et al. \cite{Schomaker:2015}.

For this analysis, we are particularly interested in the cumulative mortality difference between strategies (i) `immediate ART initiation' and (ii) `assign ART if CD4 count $<350$ cells/mm$^3$ or CD4\% $<15\%$', i.e. we are comparing current practices with those in place in 2006. We can estimate these quantities using the g-formula, see Appendix A for a comprehensive summary of our implementation details and assumptions. The standard way to obtain 95\% confidence intervals for this method is using bootstrapping. However, baseline data of CD4 count, CD4\%, HAZ, and WAZ are missing: 18\%, 28\%, 40\%, and 25\% respectively. We use multiple imputation (using \texttt{Amelia II} \cite{Honaker:2011}) to impute this data. We also impute follow-up data after nine months without any visit data, as from there on it is plausible that follow-up measurements that determine ART assignment (e.g. CD4 count) were taken (and are thus needed to adjust for time-dependent confounding) but were not electronically recorded, probably because of clerical and administrative errors. Under different assumptions imputation may not be needed. To combine the $M=10$ imputed data sets with bootstrap estimation ($B=200$) we use the four approaches introduced in Section \ref{sec:methods}: MI Boot, MI Boot (PS), Boot MI, and Boot MI (PS).

Three year mortality for immediate ART initiation was estimated as 6.08\%, whereas mortality for strategy (ii) was estimated as 6.87\%. This implies a mortality difference of 0.79\%. The results of the respective confidence intervals are summarized in Figure 3: the estimated mortality differences are $[-0.34\%;1.61\%]$ for Boot MI (PS), $[0.12\%;1.07\%]$ for Boot MI, $[-0.31\%;1.63\%]$ for MI Boot (PS), and $[-0.81\%;2.40\%]$ for MI Boot.

Figure 3 shows that the confidence intervals vary with respect to the different approaches: the shortest interval is produced by the method Boot MI. Note that only for this method the 95\% confidence interval does not contain the $0$\% when estimating the mortality difference, and therefore suggests a beneficial effect of immediate treatment initiation. The distributions of $\hat{\theta}^{\ast}_{b,m}$ for Boot MI (PS) and MI Boot (PS), as well as the distribution of $\hat{\bar\theta}^{\ast}_{b}$ for Boot MI, are also visualized in the figure and are reasonably symmetric.

Figure 4 visualizes both the bootstrap distributions in each imputed data set (method MI Boot [PS]) as well as the distribution of the estimators in each bootstrap sample (method Boot MI [PS]). It is evident that the overall variation of the estimates is similar for these two approaches considered, which explains why their confidence intervals in Figure 3 are almost identical. Moreover, and of note, the top panel highlights the large variability of the point estimates used for the calculation of the MI Boot estimator. The graph indicates a large between imputation uncertainty of the point estimates, possibly due to the high missingness and complex imputation procedure. The large confidence interval of MI Boot in Figure 3, based on formula (\ref{formula_MI_var}), reflects this uncertainty.

\begin{figure}[h!]
\begin{center}
\includegraphics[scale=0.65]{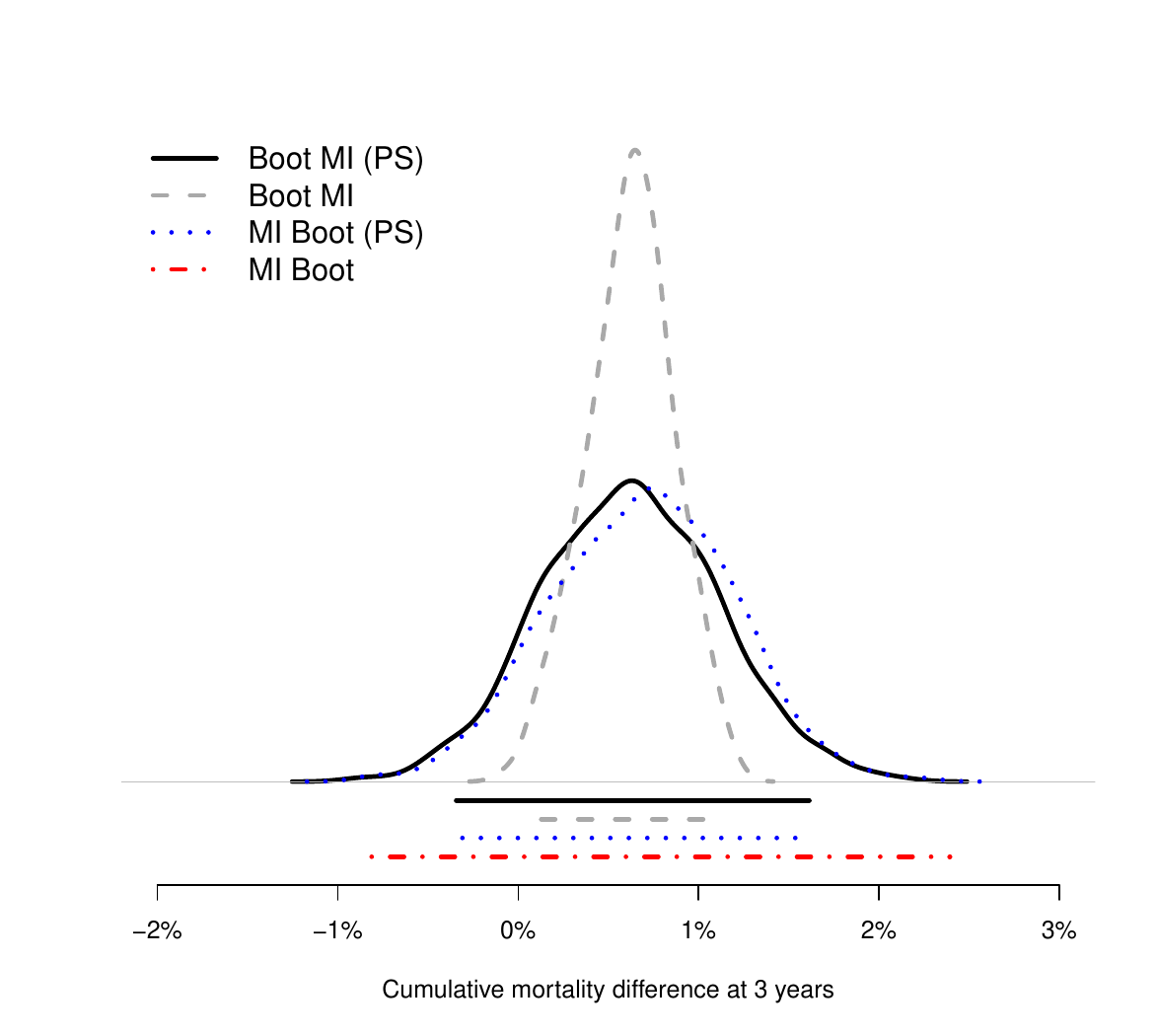}
\caption{\label{figure:sim2}Estimated cumulative mortality difference between the interventions `immediate ART' and `350/15' at 3 years: distributions and confidence intervals of different estimators}
\end{center}
\end{figure}

\begin{figure}[h!]
\begin{center}
\includegraphics[scale=0.65]{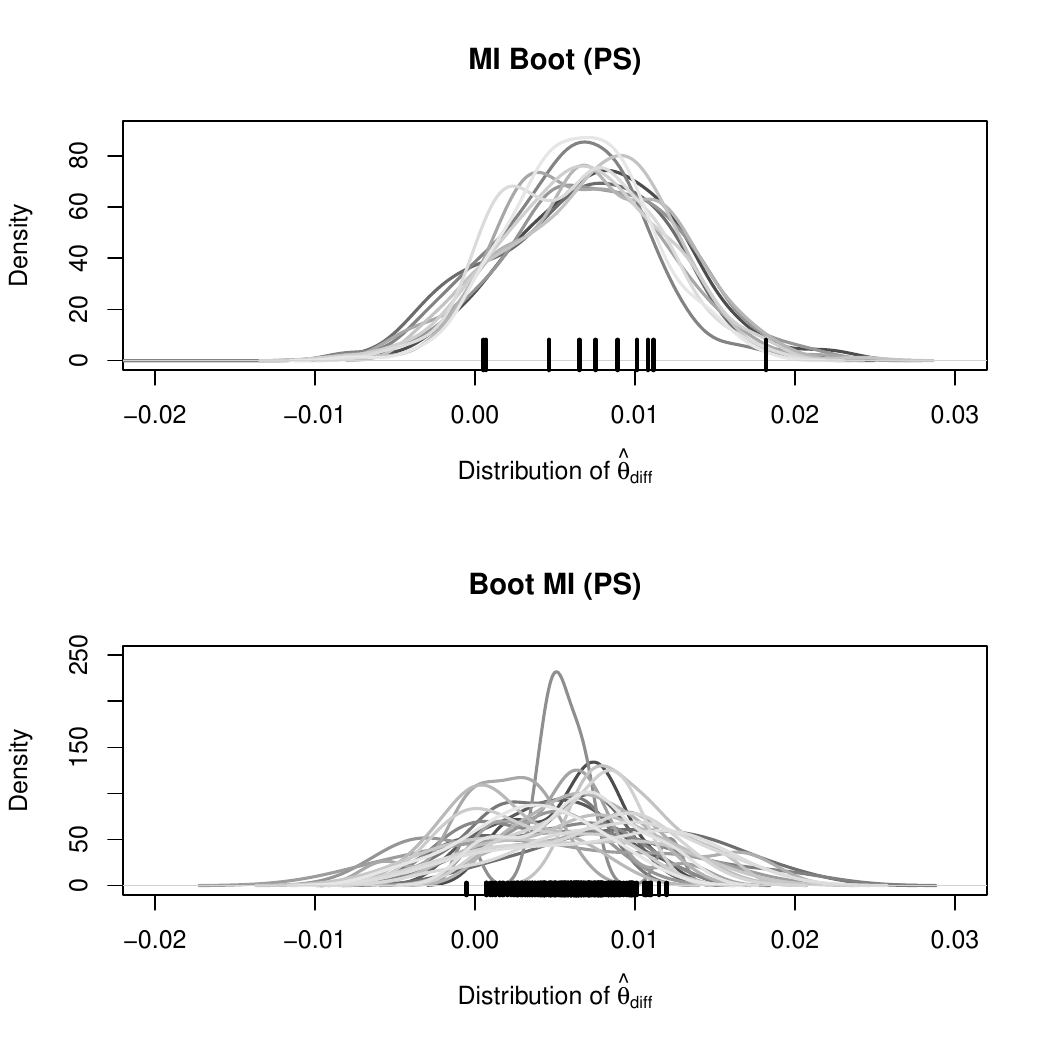}
\caption{\label{figure:data1}Estimated cumulative mortality difference: distribution of `MI Boot (PS)' for each imputed dataset (top) and distribution of `Boot MI (PS)' for 25 random bootstrap samples (bottom). Point estimates are marked by the black tick marks on the x-axis.}
\end{center}
\end{figure}

In summary, the above analyses suggest a beneficial effect of immediate ART initiation compared to delaying ART until CD4 count $<350$ cells/mm$^3$ or CD4\% $<15\%$ when using method 3, Boot MI. The other methods produce larger confidence intervals and do not necessarily suggest a clear mortality difference.

\section{Theoretical Considerations}\label{sec:theory}
For the purpose of inference we are interested in the observed data posterior distribution of $\theta|\mathcal{D}_{obs}$ which is
\begin{eqnarray}\label{eqn:MI_integral}
    P(\theta|\mathcal{D}_{obs}) &=& \int P(\theta|\mathcal{D}_{obs},\mathcal{D}_{mis}) P(\mathcal{D}_{mis}|\mathcal{D}_{obs}) d\mathcal{D}_{mis}\nonumber\\
    &=& \int P(\theta|\mathcal{D}_{obs},\mathcal{D}_{mis}) \left\{ \int P(\mathcal{D}_{mis}|\mathcal{D}_{obs},\vartheta) P(\vartheta|\mathcal{D}_{obs}) d\vartheta \right\} \ d\mathcal{D}_{mis}\,.
    \end{eqnarray}
Please note that $\vartheta$ refers to the parameters of the imputation model whereas $\theta$ is the quantity of interest from the analysis model. With multiple imputation we effectively approximate the integral (\ref{eqn:MI_integral}) by using the average
\begin{eqnarray}\label{eqn:MI_average}
P(\theta|\mathcal{D}_{obs}) &\approx& \frac{1}{M} \sum_{m=1}^M P(\theta|\mathcal{D}_{mis}^{(m)},\mathcal{D}_{obs})
\end{eqnarray}
where $\mathcal{D}_{mis}^{(m)}$ refers to draws (imputations) from the posterior predictive distribution $P(\mathcal{D}_{mis}|\mathcal{D}_{obs})$.

\subsubsection*{MI Boot and MI Boot (PS)}
The MI Boot method essentially uses rules (\ref{formula_MI_point})  and (\ref{formula_MI_var}) for inference, where, for a given scalar, the respective variance in each imputed data set $\widehat{{\text{Var}}}(\hat{\theta}_m)$ is not calculated analytically but using bootstrapping. This approach will work if the bootstrap variance for the imputed data set is close to the analytical variance. If there is no analytical variance, it all depends on various factors such as sample size, estimator of interest, proportion of missing data, and others. The data example highlights that in complex settings with a lot of missing data the between imputation variance can be large, yielding conservative interval estimates. As well-known from MI theory $M$ should, in many settings, be much larger than 5 for good estimates of the variance \cite{Little:2002}. Using bootstrapping to estimate the variance does not alter these conclusions. Using MI Boot should always be complemented with a reasonably large number of imputations. This consideration also applies to MI Boot pooled, which --as seen in the simulations--, can sometimes be even more sensitive to the choice of $M$.

\subsubsection*{Boot MI and Boot MI (PS)}
Boot MI uses $\mathcal{D}=\{\mathcal{D}_{mis},\mathcal{D}_{obs}\}$ for bootstrapping. Most importantly, we estimate $\theta$, the quantity of interest, in each bootstrap sample using multiple imputation. We therefore approximate $P(\theta|D_{obs})$ through (\ref{eqn:MI_integral}) by using multiple imputation to obtain $\hat{\theta}$ and bootstrapping to estimate its distribution --  which is valid under the missing at random assumption.

However, if we simply pool the data and apply the method Boot MI (PS) we essentially pool all estimates $\hat{\theta}_{m,b}$: with this approach each of the $B \times M$ estimates $\hat{\theta}_{m,b}$ serves then as an estimator of $\theta$ (as we do not combine/average any of them). A possible interpretation of this observation is that each $\hat{\theta}_{m,b}$ estimates $\theta$ and since this is only a single draw from the posterior predictive distribution $P(\mathcal{D}_{mis}|\mathcal{D}_{obs})$ we conduct multiple imputation with $M=1$, i.e. we calculate $\hat{\bar\theta}_{\text{MI}} = \frac{1}{1} \sum_{m=1}^{1} \hat{\theta}_{m,b}$,  $B \times M$ times. Such an estimator is statistically inefficient as we know from MI theory: the relative efficiency of an MI based estimator (compared to the true variance) is $(1+\frac{\gamma}{M})^{-1}$ where $\gamma$ describes the fraction of missingness (i.e. $V/(W+V)$) in the data. For example, if the fraction of missingness is $0.25$, and $M=5$, then the loss of efficiency is 5\% \cite{White:2011}. The lower $M$, the lower the efficiency, and thus the higher the variance. This explains the results of the simulation studies: pooling the estimates is inefficient, does therefore overestimate the variance, and thus leads to confidence intervals with incorrect coverage.

It follows that one typically gets larger interval estimates when using Boot MI (PS) instead of Boot MI. Similarly, one can decide to use Boot MI with $M=1$, which is not incorrect but often inefficient in terms of interval estimation.

\subsubsection*{Comparison}
General comparisons between MI Boot and Boot MI are difficult because the within and between imputation uncertainty, as well as the within and between bootstrap sampling uncertainty, will determine the actual width of a confidence interval. If the between imputation uncertainty is large compared to between bootstrap sample uncertainty (as, for example, in the data example [Figure 4]) then MI Boot is large compared to Boot MI. However, if the between imputation uncertainty is small relative to the bootstrap sampling uncertainty, then Boot MI may give a similar confidence interval to MI Boot (as in the simulations [Figure 2]).

Another consideration is related to the application of the bootstrap. We have focused on the percentile method to create confidence intervals. However, it is also possible to create bootstrap intervals based on the $t-$distribution. Here, an estimator's variance is estimated with the sample variance from the $B$ bootstrap estimates and symmetric confidence intervals are generated based on an appropriate $t$-distribution. In fact, MI Boot uses this approach because in each imputed dataset we estimate the bootstrap variance $\widehat{\text{Var}}(\hat{\theta}_m)$ $=$ $ (B-1)^{-1} \sum_b (\hat{\theta}_{m,b} - \hat{\bar{\theta}}_{m,b})^2$, then calculate (\ref{formula_MI_var}), followed by confidence intervals based on a $t_{R}$ distribution, see Section \ref{sec:methods}. A similar approach would be possible when applying Boot MI. This method produces $B$ point estimates $\hat{\bar{\theta}}^{\ast}_b = M^{-1} \sum_m \hat{\theta}^{\ast}_{b,m}$ for $\theta$. One could estimate the variance as $(B-1)^{-1} \sum_b (\hat{\bar{\theta}}^{\ast}_b - \hat{\bar{\bar{\theta}}}^{\ast})^2$, with $\hat{\bar{\bar{\theta}}}^{\ast} = B^{-1} \sum_b \hat{\bar{\theta}}^{\ast}_b$, and then create confidence intervals based on a $t$-distribution. This would however require that one assumes the estimator to be approximately normally distributed.

\subsubsection*{Bootstrapping as part of the imputation procedure}
For each of the estimators introduced in Section \ref{sec:methods}, $M$ \textit{proper} multiply imputed data sets are needed. ``Proper'' means that the application of formulae (\ref{formula_MI_point}) and (\ref{formula_MI_var}) yield 1) approximately unbiased point estimates and 2) interval estimates which are randomization valid in the sense that actual interval coverage equals the nominal interval coverage. Some imputation algorithms use bootstrapping to create proper imputations, and this may not be confused with the bootstrapping step \textit{after} multiple imputation which we focus on in this paper.

To follow this argument in more detail it is important to understand that proper imputations are created by means of random draws from the posterior predictive distribution of the missing data given the observed data (or an approximation thereof). These draws can (i) either be generated by specifying a multivariate distribution of the data (joint modeling) and simulate the posterior predictive distribution with a suitable algorithm; or (ii) by specifying individual conditional distributions for each variable $\mathbf{X}_j$ given the other variables (fully conditional modeling)  and iteratively drawing and updating imputed values from these distributions which will then (ideally) converge to draws of the theoretical joint distribution; or (iii) by the use of  alternative algorithms.

An example for (i) is the EMB algorithm from the $R$-package \texttt{Amelia II} which assumes a multivariate normal distribution for the data, $\mathcal{D} \sim N(\mu,\Sigma)$ (possibly after suitable transformations beforehand). Then, $B$ bootstrap samples of the data (including missing values) are drawn and in each bootstrap sample the EM algorithm \cite{Dempster:1977} is applied to obtain estimates of $\mu$ and $\Sigma$ which can then be used to generate proper multiple imputations by means of the sweep-operator \cite{Goodnight:1979, Honaker:2010}. Of note, the algorithm can handle highly skewed variables by imposing transformations on variables (log, square root). Categorical variables are recoded into dummy variables based on the knowledge that for binary variables the multivariate normal assumption can yield good results \cite{Schafer:2002}.

An example for (ii) is imputation by chained equations (ICE, mice). Here, (a) one first specifies \textit{individual} conditional distributions (i.e. regression models) $p(\mathbf{X}_j|\mathbf{X}_{-j},\theta_j)$ for each variable. Then, (b) one iteratively fits  all regression models and generates \textit{random} draws of the coefficients, e.g. $\tilde{\beta}\sim N(\hat{\beta},\widehat{\text{Cov}}(\hat{\beta}))$. Values are (c) imputed as random draws from the distribution of the regression predictions. Then, (b) and (c) are repeated $k$ times until convergence. The process of iteratively drawing and updating the imputed values from the conditional distributions can be viewed as a Gibbs sampler that converges to draws from the (theoretical) joint distribution. This method is among the most popular ones in practice and has been implemented in many software packages \cite{vanBuuren:2011, Royston:2011}. However, there remain theoretical concerns as a joint distribution may not always exist for a given specifications of the conditional distributions \cite{Drechsler:2008}. A variation of (c) is a fully Bayesian approach where the posterior predictive distribution is used to draw imputations. Here, the bootstrap is used to model the imputation uncertainty and to draw the $M$ imputations needed for the $M$ imputed data sets. This variation yields approximate proper imputations and is implemented in the $R$ library \texttt{Hmisc} \cite{Harrell:2016}.

An example for (iii) is the Approximate Bayesian Bootstrap \cite{Rubin:1986}. Here, the (cross-sectional) data is stratified into several strata, possibly by means of the covariates of the analysis model. Then, within each stratum (a) one draws a bootstrap sample among the complete data (with respect to the variable to be imputed). Secondly, (b) one uses the original data set (with missing values) and imputes the missing data based on units from the data set created in (a), with equal selection probability and with replacement. The multiply imputed data are obtained by repeating (a) and (b) $M$ times.

It is evident from the above examples that many imputation methods use bootstrap methodology as part of the imputation model, that this does not replace the additional bootstrap step needed for the inference in the analysis model, and that -- if they are combined -- the resampling steps are nested.

\section{Conclusion}\label{sec:conclusion}
The current statistical literature is not clear on how to combine bootstrap with multiple imputation inference. We have proposed that a number of approaches are intuitively appealing and three of them are correct: Boot MI, MI Boot, MI Boot (PS). Using Boot MI (PS) can lead to too large and invalid confidence intervals and is therefore not recommended.

Both Boot MI and MI Boot are probably the best options to calculate randomization valid confidence intervals when combining bootstrapping with multiple imputation. As a rule of thumb, our analyses suggest that the former may be preferred for small $M$ or large imputation uncertainty and the latter for normal $M$ and little/normal imputation uncertainty.

There are however other considerations when deciding between MI Boot and Boot MI. The latter is computationally much more intensive. This matters particularly when estimating the analysis model is simple in relation to creating the imputations. In fact, in our first simulation this affected the computation time by a factor of 13.  However, MI Boot naturally provides symmetrical confidence intervals. These intervals may not be wanted if an estimator's distribution is suspected to be non-normal.

\appendix

\section{Details of the G-formula Implementation}\label{sec:appendix_g_formula}

We consider $n$ children studied at baseline ($t=0$) and during discrete follow-up times ($t=1,\ldots,T$). The data consists of the outcome $Y_t$, an intervention variable $A_t$, $q$ time-dependent covariates $\mathbf{L}_t=\{L^1_t,\ldots,L^q_t\}$, and a censoring indicator $C_t$. The covariates may also include baseline variables $V=\{L^1_0,\ldots,L^{q_V}_0\}$. The treatment and covariate history of an individual $i$ up to and including time $t$ is represented as $\bar{A}_{t,i}=(A_{0,i},\ldots,A_{t,i})$ and $\bar{L}^s_{t,i}=(L^s_{0,i},\ldots,L^s_{t,i})$ respectively. $C_t$ equals $1$ if a subject gets censored in the interval $(t-1,t]$, and $0$ otherwise. Therefore, $\bar{C}_t=0$ is the event that an individual remains uncensored until time $t$.

The counterfactual outcome $Y^{\bar{a}_{t}} = Y_t^{\bar{a}_t}$ refers to the hypothetical outcome that would have been observed at time $t$ if every subject had received, likely contrary to the fact, the treatment history $\bar{A}_t=\bar{a}_t$. Similarly, $\mathbf{L}_t^{\bar{a}_t}$ are the counterfactual covariates related to the intervention $\bar{A}_t=\bar{a}_t$. The above notation refers to \textit{static} treatment rules; a treatment rule may however depend on covariates, and in this case it is called \textit{dynamic}. A dynamic rule $\bar{d}(\bar{a}_{t,i}; \mathbf{\bar{L}}_{t,i})$ assigns treatment ${A}_{t,i} \in \{0,1\}$ as a function of the covariate history $\mathbf{\bar{L}}_{t,i}$ and the intervention vector $\bar{a}_{t,i}$ may therefore vary by subject $i$. The counterfactual outcome related to a dynamic rule $\bar{d}$ is $Y_{t,i}^{\bar{d}(\bar{a}_{t,i}; \mathbf{L}_{t,i})} = Y_{t,i}^{\bar{d}}$, and the counterfactual covariates are $\mathbf{L}_{t,i}^{\bar{d}}$. Often $\mathbf{\bar{A}}_{t,i} = (\bar{A}_{t,i},\bar{C}_{t,i}=0)$ which means that one is interested in the counterfactuals for intervention $\bar{A}_{t,i}$ under (the intervention of) no censoring. In our notation, for simplicity, a rule $\bar{d}$ can be dynamic and intervene on multiple variables, including the censoring mechanism, without referring to it explicitly, i.e. $\bar{d}$ may relate to $\bar{d}(\bar{a}_{t,i}, \bar{c}_{t,i}; \mathbf{\bar{L}}_{t,i})$. We write $\bar{a}_{t,i}^{\bar{d}}$ for the intervention history individual $i$ received under rule $\bar{d}$.

In our setting we study $n=5826$ children for $t=0,1,3,6,9,\ldots$ where the follow-up time points refer to the intervals $(0, 1.5)$, $[1.5, 4.5)$, $[4.5, 7.5)$, $\ldots$, $[28.5, 31.5)$, $[31.5, 36)$ months respectively. Follow-up measurements, if available, refer to measurements closest to the middle of the interval. In our data ${Y}_t$ refers to death at time $t$ (i.e. occurring during the interval $(t-1,t]$). ${A}_t$ refers to antiretroviral treatment (ART) taken at time $t$. $\mathbf{L}_t=(L^1_t, L_t^2, L^3_t, L^{1m}_t, L_t^{2m}, L^{3m}_t)$ are CD4 count, CD4\%, and weight for age z-score (WAZ, which serves as a proxy for WHO stage, see \cite{Schomaker:2013} for more details) as well as three indicator variables whether these variables have been measured at time $t$ or not. $\mathbf{V} = \mathbf{L}_0^V$ refer to  baseline values of CD4 count, CD4\%, WAZ, height for age z-score (HAZ) as well as sex, age, and region. The two treatment rules of interest are:\vspace*{-0.5cm}

\begin{flalign*}
\bar{d}_{t,i}^{1} &= \,\, \left\{ c_{t,i} = 0; \quad l^{1m}_{t,i} = l^{2m}_{t,i} = l^{3m}_{t,i} = 1; \quad a_{t,i}=1 \quad \text{for} \quad \forall t,i\right.& \\
\bar{d}_{t,i}^{2} &= \left\{ \begin{array}{cl}
               c_{t,i} = 0; \quad l^{1m}_{t,i} = l^{2m}_{t,i} = l^{3m}_{t,i} = 1; \quad a_{t,i}=1 & \quad \mbox{if} \quad \text{CD4 count}_{t,i}^{\bar{d}} < 350  \quad \text{or} \quad \text{CD4\%}^{\bar{d}}_{t,i} < 15\%\\
               c_{t,i} = 0; \quad l^{1m}_{t,i} = l^{2m}_{t,i} = l^{3m}_{t,i} = 1; \quad a_{t,i}=0 &  \quad \mbox{otherwise}
               \end{array}
               \right. &
\end{flalign*}

The quantity of interest is thus cumulative mortality after $T=36$ months, under (the intervention of) no censoring, regular 3 monthly follow-up and for treatment assignment according to $\bar{d}_j$, that is $\psi= \sum_{t=1}^{T} \mathbb{P}(Y_t^{\bar{d}}=1)$.

Under the assumption of \textit{consistency}, i.e. $Y^{\bar{d}} = Y$ if $\bar{A}_t = \bar{a}_{t,i}^{\bar{d}}$ and $\bar{\mathbf{L}}_t^{\bar{d}}=\bar{\mathbf{L}}_t$ if $\bar{A}_{t-1} = \bar{a}_{t-1,i}^{\bar{d}}$, \textit{sequential conditional exchangeability} (or \textit{no unmeasured confounding}), i.e. $Y^{\bar{d}} \coprod {A_t|\bar{L}_t, \bar{A}_{t-1}}$ for $\forall \bar{A}_t=\bar{a}_{t}^{\bar{d}}, \bar{L}_t=\bar{l}_t, t \in \left\{0,\ldots,T\right\}$ and \textit{positivity}, i.e. $P(A_t=\bar{a}_{t}^{\bar{d}}|\bar{\mathbf{L}}_t=\bar{\mathbf{l}}_t,\bar{A}_{t-1}=\bar{a}_{t-1}^{\bar{d}})>0$ for $\forall t,\bar{a}_t^{\bar{d}},\bar{\mathbf{l}}_t$ with $P(\bar{\mathbf{L}}_t=\bar{l}_t,\bar{A}_{t-1}=\bar{a}_{t-1}^{\bar{d}})\neq 0$, the g-computation formula can estimate $\psi$ as:

\begin{eqnarray}\label{eqn:gformula}
\hspace*{1cm}\psi= \sum_{t=1}^{T} \mathbb{P}(Y_t^{\bar{d}}=1) &= & \sum_{t=1}^T \int_{\mathbf{\bar{l}}\in\mathbf{\bar{L}}_t}
\left\{
\begin{aligned}
&\mathbb{P}(Y_t=1|\bar{A}_{t-1}=\bar{a}_{t-1,i}^{\bar{d}}, \mathbf{\bar{L}}_t=\mathbf{\bar{l}}_t, Y_{t-1}=0
)\times\\
&\prod_{t=1}^{T}
  f(\mathbf{L}_t|\bar{A}_{t-1}=\bar{a}_{t-1,i}^{\bar{d}}, \mathbf{\bar{L}}_{t-1}=\mathbf{\bar{l}}_{t-1}, Y_{t-1}=0)
\end{aligned}
\right\}
\, d\mathbf{\bar{l}}
\end{eqnarray}
see \cite{Westreich:2012} and \cite{Young:2011} about more details and implications of the representation of the g-formula in this context. Note that the inner product of (\ref{eqn:gformula}) can be written as
\begin{eqnarray*}\label{eqn:gformula2}
\prod_{t=1}^{T} \prod_{s=1}^{q}
f({L}_t^s|\bar{A}_{t-1}\hspace*{-0.1cm}=\bar{a}_{t-1,i}^{\bar{d}}, \mathbf{\bar{L}}_{t-1}\hspace*{-0.1cm}=\mathbf{\bar{l}}_{t-1}, \mathbf{L}_t^1 \hspace*{-0.1cm}= \mathbf{l}_t^1,\ldots, L_t^{s-1}\hspace*{-0.1cm}=l_t^{s-1}, \bar{Y}_{t-1}\hspace*{-0.1cm}=0)\,.
\end{eqnarray*}
In the above representation of the g-formula we assume that the time ordering is $L^1_t \rightarrow L^2_t \rightarrow L^3_t \rightarrow A/C \rightarrow Y$.

There is no closed form solution to estimate (\ref{eqn:gformula}), but $\theta$ can be approximated by means of the following algorithm; Step 1: use additive linear and logistic regression models to estimate the conditional densities on the right hand side of (\ref{eqn:gformula}), i.e. fit regression models for the outcome variables CD4 count, CD4\%, WAZ, and death at $t=1,3,..,36$ using the available covariate history and model selection. Step 2: use the models fitted in step 1 to stochastically generate $\mathbf{L}_t$ and $Y_t$ under a specific treatment rule. For example, for rule (ii), draw $\mathbf{L}_1^1 = \sqrt{\text{CD4 count}_1}$ from a normal distribution related to the respective additive linear model from step 1 using the relevant covariate history data. Set $A_1 = 1$ if the generated CD4 count at time 1 is $<350$ cells/mm$^3$ or CD4\% $<15$\% (for rule $\bar{d}^2$). Use the simulated covariate data and treatment as assigned by the rule to generate the full simulated data set forward in time and evaluate cumulative mortality after 3 years of follow-up. We refer the reader to \cite{Schomaker:2015}, \cite{Westreich:2012}, and \cite{Young:2011} to learn more about the g-formula in this context.

Note that the so-called sequential g-formula, used in the simulation study, shares the idea of standardization in the sense that one sequentially marginalizes the distribution with respect to $\mathbf{L}$ given the intervention rule of interest. It is just a re-expression of (\ref{eqn:gformula}) where integration with respect $\mathbf{L}$ is not needed \cite{Petersen:2014a}:

\begin{eqnarray}\label{eqn:seq_g_formula}
\hspace*{1cm}\mathbb{E}(Y_T^{\bar{d}}) &=& \mathbb{E}(\,\mathbb{E}(\,\ldots\mathbb{E}(\,\mathbb{E}(Y_T|\bar{A}_T=\bar{a}_{T}^{\bar{d}}, \mathbf{\bar{L}}_T) | \bar{A}_{T-1}=\bar{a}_{T-1}^{\bar{d}}, \mathbf{\bar{L}}_{T-1}\,)\ldots|\bar{A}_{0}=\bar{a}_{0}^{\bar{d}}, \mathbf{\bar{L}}_{0}\,)|\mathbf{\bar{L}}_{0}\,)\,.
\end{eqnarray}

\section{Data Generating Process in the Simulation Study}\label{sec:appendix_data_generating}
\begin{sidewaysfigure}[h!]
\begin{center}
\includegraphics[scale=1, angle=270, trim=4cm 3cm 4cm 3cm]{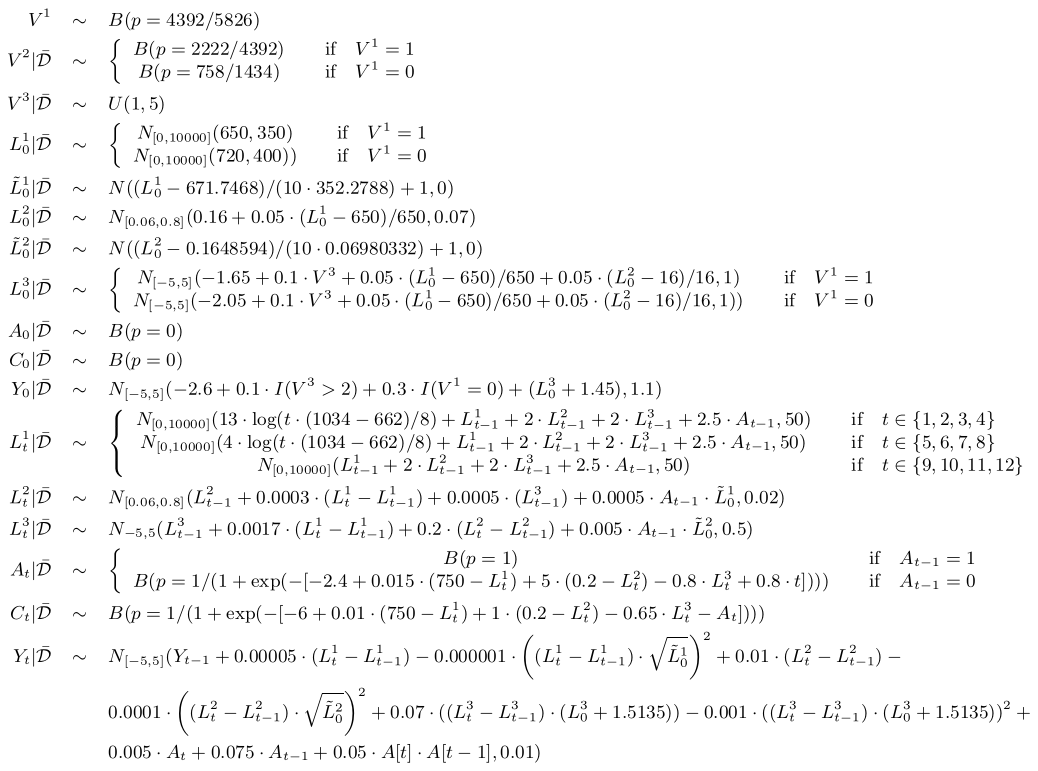}
\caption{Data generating process in the simulation study}
\label{figure:data_generating}
\end{center}
\end{sidewaysfigure}

Both baseline data ($t=0$) and follow-up data ($t=1,\ldots,12$) were created using structural equations using the $R$-package \texttt{simcausal} \cite{Sofrygin:2016}. The below listed distributions, listed in temporal order, describe the data-generating process. Baseline data refers to region, sex, age, CD4 count, CD4\%, WAZ and HAZ respectively ($V^1$, $V^2$, $V^3$, $L^1_0$, $L^2_0$, $L^3_0$, $Y_0$). Follow-up data refers to CD4 count, CD4\%, WAZ and HAZ ($L^1_t$, $L^2_t$, $L^3_t$, $Y_t$), as well as an antiretroviral treatment ($A_t$) and censoring ($C_t$) indicator. For simplicity, no deaths are assumed. In addition to Bernoulli ($B$), uniform ($U$) and normal ($N$) distributions, we also use truncated normal distributions which are denoted by $N_{[a,b]}$ where $a$ and $b$ are the truncation levels. Values which are smaller $a$ are replaced by a random draw from a $U(a_1,a_2)$ distribution and values greater than $b$ are drawn from a $U(b_1,b_2)$ distribution. Values for $(a_1, a_2, b_1, b_2)$ are $(0,50,5000,10000)$ for $L^1$, $(0.03,0.09,0.7,0.8)$ for $L^2$, and $(-10,3,3,10)$ for both $L^3$ and $Y$. The notation $\bar{\mathcal{D}}$ means ``conditional on the data that has already been measured (generated) according the the time ordering''. The distributions are listed in Figure \ref{figure:data_generating}

The data generating process leads to the following baseline values: region A = 75.5\%; male sex = 51.2\%; mean age  = 3.0 years; mean CD4 count = 672.5; mean CD4\% = 15.5\%; mean WAZ = -1.5; mean HAZ = -2.5. At $t=12$ the arithmetic mean of CD4 count, CD4\%, WAZ and HAZ are 1092, 27.2\%, -0.8, -1.5 respectively. The target quantities $\psi_1$ and $\psi_2$ are defined as the expected value of $Y$ at time $T$, under no censoring, for a given treatment rule $\bar{d}^j$, where
\begin{eqnarray*}
\bar{d}_{t,i}^{1} = \,\, \left\{ c_{t,i} = 0; \quad a_{t,i}=1 \quad \text{for} \quad \forall t,i\right. \quad\quad \text{and}  \quad\quad
\bar{d}_{t,i}^{2} = \,\, \left\{ c_{t,i} = 0; \quad a_{t,i}=0 \quad \text{for} \quad \forall t,i\right.
\end{eqnarray*}
and are $-1.03$ and $-2.45$ respectively. Missing baseline and follow-up data were created based on the following functions:
\begin{eqnarray*}
\pi_{(L^1_t)} &=& 0.1;\\
\pi_{(L^2_0)}(L^1_0) &=& 1 - \frac{1}{(0.001\cdot L^1_0)^2+1}\, ;\quad\quad
\pi_{(L^2_t)}(t,L^1_t) = 1 - \frac{1}{(0.00005 \cdot t \cdot L^1_t)^2+1} ;\\
\pi_{(L^3_0)}(Y_0) &=& 1 - \frac{1}{(0.2\cdot |Y_0|)^2+1}\, ;\quad\quad
\pi_{(L^3_t)}(t,Y_t) = 1 - \frac{1}{(0.015 \cdot t \cdot |Y_t|)^2+1} ;\\
\pi_{(Y_0)}(L^3_0) &=& 1 - \frac{1}{(0.7\cdot |L^3_0|)^2+1}\, ;\quad\quad
\pi_{(Y_t)}(t,L^3_t) = 1 - \frac{1}{(0.015 \cdot t \cdot |L^3_t|)^2+1} \,.\\
\end{eqnarray*}

\clearpage
\section*{Acknowledgements}
The authors gratefully acknowledge Mary-Ann Davies and Valeriane Leroy who contributed to the analysis and study design of the data analysis. We further thank Lorna Renner, Shobna Sawry, Sylvie N'Gbeche, Karl-G{\"u}nter Technau, Francois Eboua, Frank Tanser, Haby Syg\-nate-Sy, Sam Phiri, Madeleine Amorissani-Folquet, Vivian Cox, Fla Koueta, Cleophas Chimbete, Annette Lawson-Evi, Janet Giddy, Clarisse Amani-Bosse, and Robin Wood for sharing their data with us. We would also like to highlight the support of the Pediatric West African Group and the Paediatric Working Group Southern Africa. The NIH has supported the above individuals, grant numbers 5U01AI069924-05 and U01AI069919. We also thank Jonathan Bartlett for his feedback on an earlier version of this paper.

\bibliographystyle{unsrt}
\bibliography{MIandB_revision1}
\end{document}